\documentclass[aps, prb, amsmath,amssymb, reprint, showpacs,superscriptaddress,groupedaddress, nofootinbib ]{revtex4-1}

\usepackage{nicefrac}
\usepackage{float}
\usepackage{graphicx}
\usepackage{dcolumn}
\usepackage{bm}
\usepackage{comment}
\usepackage{soul}
\usepackage{hyperref}
\hypersetup{
	colorlinks=true,
	citecolor=blue,
	linkcolor=blue,
	filecolor=blue,      
	urlcolor=blue,
}

\begin{document}

\title{Disorder-mediated quenching of magnetization in NbVTiAl: Theory and Experiment}

\author{Deepika Rani,$^{1,2}$ Jiban Kangsabanik,$^{1,3}$ K. G. Suresh$^{1}$ and Aftab Alam$^{1}$}
\affiliation{$^1$Department of Physics, Indian Institute of Technology Bombay, Powai, Mumbai 400076, Maharashtra, India\\
$^2$Department of Physics, Indian Institute of Technology Delhi, Hauz Khas, New Delhi-110016, India\\
$^3$Department of Physics, Technical University of Denmark (DTU), Lyngby 2800 Kgs, Denmark}
\date{\today}

\begin{abstract}
In this paper, we present the structural, electronic, magnetic and transport properties of a equiatomic quaternary alloy NbVTiAl. The absence of (111) and (200) peaks in X-ray diffraction (XRD) data confirms the A2-type structure. Magnetization measurements indicate a high Curie temperature and a negligibly small magnetic moment ($\sim 10^{-3} \mu_B/f.u.$) These observations are indicative of fully compensated ferrimagnetism in the alloy. Temperature dependent resistivity indicate metallic nature. Ab-initio calculation of fully ordered NbVTiAl structure confirms a nearly half metallic behavior with a high spin polarization ($\sim$ 90 \%) and a net magnetic moment of 0.8 $\mu_B/f.u.$ (in complete contrast to the experimental observation). One of the main objective of the present paper is to resolve and explain the long-standing discrepancy between theoretical prediction and experimental observation of magnetization for V-based quaternary Heusler alloys, in general. To gain an in-depth understanding, we modelled various disordered states and its subsequent effect on the magnetic and electronic properties. The discrepancy is attributed to the A2 disorder present in the system, as confirmed by our XRD data. The presence of disorder also causes the emergence of finite states at the Fermi level, which impacts the spin polarization of the system. 
\end{abstract}

\maketitle

\section{Introduction}
Magnetic materials with high spin polarization have always been the centre of attraction for spintronics based research. Heusler alloys are indeed one of the most promising class of such materials. They have shown immense potential for spintronics application by virtue of their tunable and diverse electronic/magnetic properties. Half-metallic ferromagnets (HMF)\cite{PhysRevLett.50.2024} and spin gapless semiconductors (SGS)\cite{WANG20201,doi:10.1063/5.0028918} hold a special place in this class of materials due to their unique band structure and high spin polarization. Due to this, intensive efforts have been put in to study the magnetic and electronic properties of these materials. First-principles calculations predict half-metallicity with 100\% spin-polarization for several pure Heusler alloys. It is, however, important to note that most of the ab-intio investigations are based on defect free or stoichiometric composition of the concerned material. On the contrary, Heusler alloys are know to be prone to disorder and when synthesized, they show either partial B2-type or A2-type structural disorders, which may destroy the half-metallicity of the concerned alloy. This fundamental discrepancy is mainly because of a lack of microscopic understanding of the structural and electronic/magnetic properties of such alloys. Thus, it is quite important to resolve the above mentioned discrepancy between theoretical and experimental findings in order to design new plausible Heusler-based materials for potential applications.

In this paper, we report a detailed study of structure, electronic, magnetic and transport properties of a equiatomic quaternary alloy NbVTiAl. Experimental findings are corroborated with the results obtained from first principles calculations. NbVTiAl is a Heusler-type alloy whose magnetic properties deviate from the Slater Pauling (SP) rule.\cite{doi:10.1063/1.4805063} Magnetic measurements yield an almost vanishing magnetization ($\sim 10^{-3} \mu_B/f.u.$), while an ordered NbVTiAl is expected to yield a net moment of 1.0 $\mu_B$/f.u. as per SP rule. Our ab-initio band structure calculation also confirms a net moment of 0.8 $\mu_B$/f.u. with a high spin polarization of 90\% (nearly half metallic) for a completely ordered NbVTiAl structure. This is a clear disagreement between the theoretical predictions and experimental results in terms of magnetic properties. To gain insight into the origin of this discrepancy, we have investigated the effect of disorder on the magnetic and electronic properties of NbVTiAl using state of art ab-intio calculations. To the best of our knowledge, the present study is one out of few, which correctly highlights the significance of disorder in explaining the correct magnetic and electronic state of a quaternary Heusler alloy within the DFT framework.

 The room temperature X-ray diffraction (XRD) data of NbVTiAl confirms an A2 type structural disorder, as confirmed by the absence of (111) and (200) superlattice reflection peaks. The quenching of magnetic moment was successfully explained by simulating the experimentally observed A2 disordered phase.  The disorder also causes a sharp decrease in the spin polarization. Magnetization measurements also reveal a high Curie temperature ($>$ 800 K). The transport measurements confirm a metallic-like behaviour, as also supported by our first principles calculation for A2 phase of NbVTiAl. Hall resistivity shows a linear dependence on the field with a positive slope. The calculated value of total density of states simulated for experimentally observed A2 structure was found to be comparable with that obtained from the heat capacity measurements. The detailed explanation in the present work sets up a stage for quaternary Heusler alloys in general, where disorder can play a decisive role in determining the correct magnetic and electronic properties and hence the reduced/quenched moment.
 
\section{Experimental details}
\label{exp}
Arc melting technique was used to prepare polycrystalline sample of NbVTiAl, using the stoichiometric amounts of high pure (at least 99.9\% purity) constituent elements in argon atmosphere. To reduce any contamination in the chamber, a Ti ingot was used as an oxygen getter. The ingots formed were flipped and melted several times for better homogeneity. Room temperature X-ray diffraction patterns were taken using Cu K$_\alpha$ radiation with the help of Paralytic X-pert diffractometer. Phase purity and crystal structure analysis were done using Fullprof suite\cite{RR}. Magnetization measurements were done using a vibrating sample magnetometer (VSM) attached to the physical property measurement system (PPMS) (Quantum design) for fields up to 40 kOe. Thermo-magnetic curves in the higher temperature range were obtained using a VSM attached with high temperature oven, under a field of 500 Oe. 
For resistivity and Hall effect measurements, polycrystalline alloys were cut into rectangular shape using a diamond wheel saw to dimensions nearly 1 mm thickness, 4.87 mm wide and length 8.62 mm. These rectangular pieces were then polished in order to further reduce the thickness (up to 0.52 mm). Electrical resistivity measurements (using dc-linear four probe method by applying a  5 mA current) have been carried out using PPMS. In this method, four contacts were made on the sample along a line. A constant current was allowed to flow through the two outermost contacts. The voltage was measured between the two inner contacts and the gap between them is taken as the sample length. In a four-probe geometry (see Fig. S2(a) in supplementary \cite{RR3}), the resistivity can be calculated as, $\rho$=RxA/l, where, R is the resistance, A is area of cross section through which the current passes (A =width x thickness), and l is the voltage lead separation ($V_+V_-$ = 5.76 mm). Hall measurements were carried out using five probe measurement, an additional module attached to the PPMS. In five-probe geometry, a fifth voltage lead is attached in parallel to one of the other voltage leads (see Fig. S2(b) in supplementary \cite{RR3}) to get an accurate value for the Hall potential. While the magnetic field is turned off, a potentiometer between the two leads is used to null the offset that is due to sample resistance. Then, once a field is applied, the measured potential drop gives only the Hall potential, plus components due to instrumental non-idealities that can be removed with AC filtering technique. 150 mA current was applied along the direction of length while the developed transverse voltage was measured using the leads connected on sides of the sample. The Hall Effect measurements were carried out in an applied magnetic field of 5 T in five probe contact geometry at different temperatures. The electrical contacts were made using silver epoxy. Specific heat measurements (by relaxation calorimetry) have been carried out using PPMS.
\section{Computational details}
\label {comp}
	First principles calculations of NbVTiAl were performed using spin polarized density functional theory (DFT)\cite{hohenberg1964inhomogeneous} as implemented within the Vienna ab initio simulation package (VASP),\cite{kresse1996efficient,kresse1996efficiency,kresse1993ab} with a projected augmented-wave (PAW) basis.\cite{kresse1999ultrasoft} Pseudopotential formalism with Perdew, Burke, and Ernzerhof (PBE) exchange-correlation functional\cite{perdew1996generalized} was used for all the calculations. A 24$\times$24$\times$24 $\Gamma$-centered K-mesh was used to perform the Brillouin zone(BZ) integration. A plane wave energy cut-off of 500 eV was used for all the calculations. All the structures are fully relaxed with total energies (forces) converged to values less than 10$^{-6}$ eV (0.01 eV/\AA). A 64 atom special quasirandom structure (SQS)\cite{zunger1990special} is generated to simulate the A2 disordered phase of NbVTiAl. SQS is a periodic structure known to mimic the random correlations accurately for a given alloy concentration. Alloy Theoretic Automated Toolkit (ATAT) \cite{van2013efficient} was used to generate the SQS structures. BZ integration for SQS case was performed using $6\times6\times6$ k-mesh.

\section{Results and Discussion}
\label{result}

\subsection{Structural Analysis}
\label{result1}
For any quaternary Heusler alloy (QHA) XX$'$YZ, there are 4 inequivalent Wyckoff positions, namely 4a, 4b, 4c and 4d. If one fix the position of Z-atom to be at 4a site, then there exists three possible non degenerate crystallographic configurations,\cite{RANI2019165662}
\begin{enumerate}
	\item Type I $\rightarrow$ X at 4d, X$'$ at 4c and Y at 4b site,
	\item Type II $\rightarrow$ X at 4b , X$'$ at 4c and Y at 4d site,
	\item	Type III $\rightarrow$ X at 4c , X$'$ at 4b and Y at 4d site
\end{enumerate}
For the configuration considering X at 4b, X$'$ at 4c, Y at 4d and Z at 4a Wyckoff positions, the structure factors for the superlattice $(111)$ and $(200)$ reflections can be written as,\cite{PhysRevB.99.104429}:\\
\begin{equation}
\mathrm{F_{111}} = 4[\mathrm{(f_Y-f_Z)-i(f_X-f_{X'})}]
\end{equation}
\begin{equation}
\mathrm{F_{200}} = 4[\mathrm{(f_Y+f_Z)-(f_X+f_{X'})}]
\end{equation}

These equations imply that, the intensity of the (111) peak should reduce or disappear in presence of B2 disorder (X and X$^\prime$ \& Y and Z atoms are randomly distributed). For a completely disordered structure i.e., A2-type (all the four atoms are randomly distributed at four Wyckoff positions), both the superlattice peaks $(111)$ and $(200)$ should be absent.

\begin{figure}[hbt]
\centering
\includegraphics[width=0.7\linewidth]{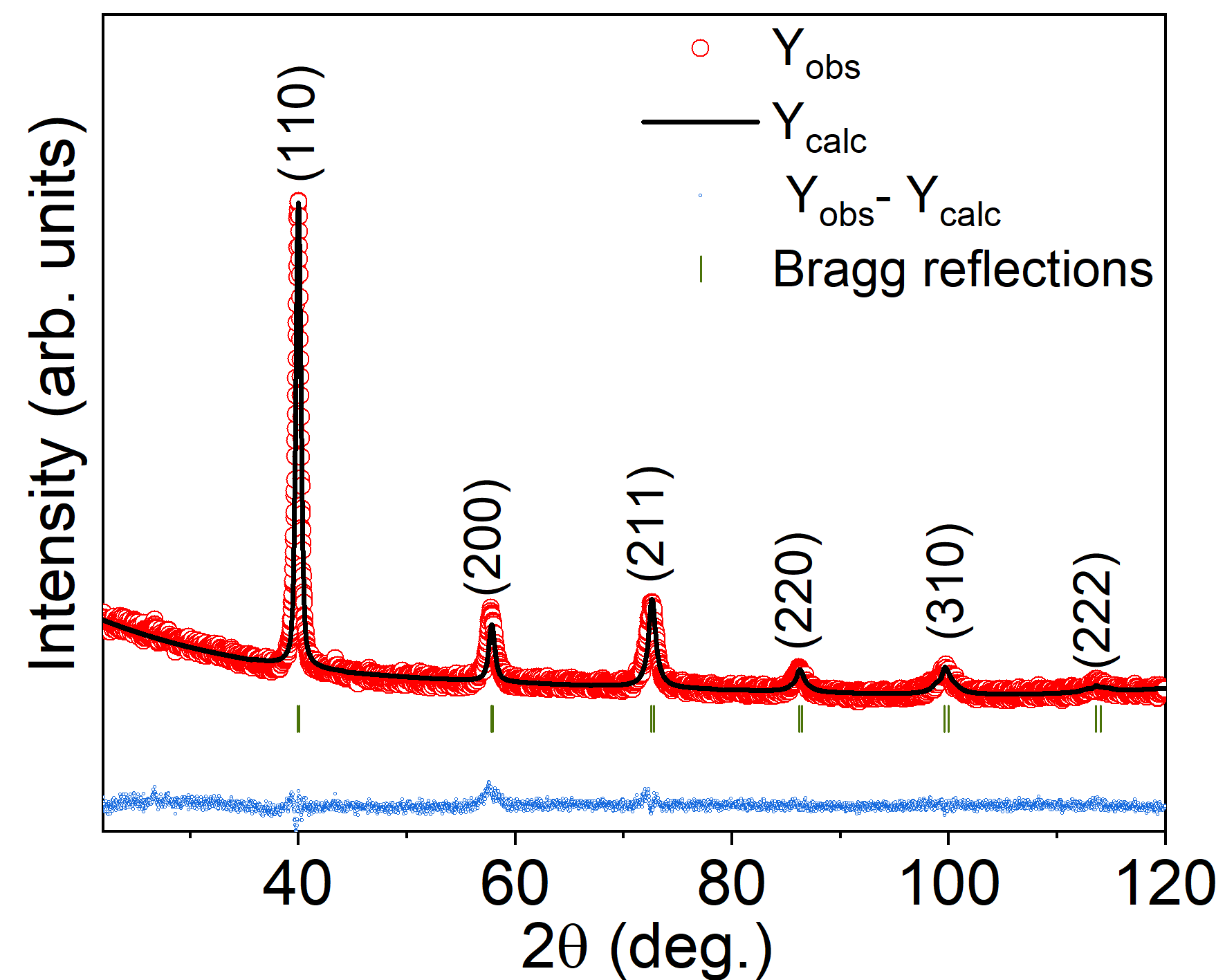}
\caption{Rietveld refined XRD pattern of NbVTiAl at room temperature. }
\label{XRD}
\end{figure}

Figure \ref{XRD} shows the rietveld refined room temperature powder XRD pattern of NbVTiAl. It is clear that the alloy exhibits a cubic crystal structure. The absence of superlattice reflection peaks (111) and (200) indicates complete A2 disorder. The lattice parameter as deduced from the refinement considering the A2-type structure (Im-3m, with all the atoms sharing 2a  site according to the stoichiometry) was found to be 3.186 $\mathrm{\AA}$. The $\chi ^2$ value was found to be 2.06. The chemical nature of the constituent elements greatly affects their ordering. In Heusler alloys, the least electronegative element occupies octahedral site.\cite{Graf20111} However, if the constituent elements have comparable electronegativities, they compete to occupy the same site, and hence cause the disorder. A possible reason of A2-type disorder in NbVTiAl could be the nearly similar electronegativities of the constituent elements (the electronegativity of Nb, V, Ti, and Al are 1.60, 1.63, 1.54, and 1.61 respectively) which make all the four sites equally probable to be occupied by the four atoms.

\begin{figure}[hbt]
	\centering
	\includegraphics[width=\linewidth]{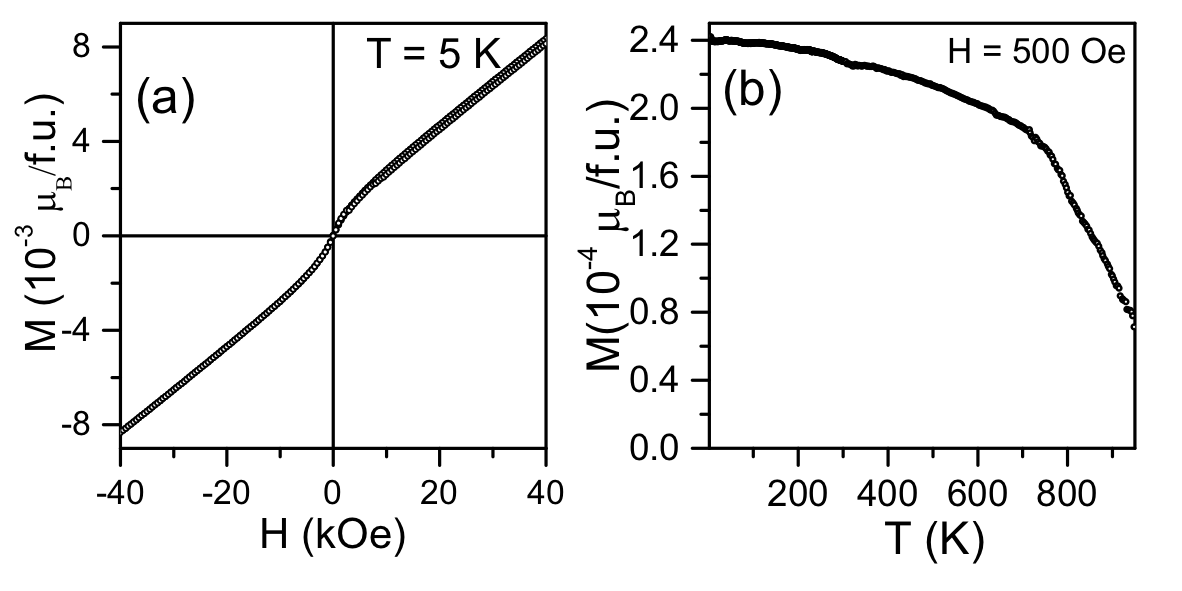}
	\caption{For NbVTiAl (a) Isothermal magnetization curve at 5 K, and (b) Temperature dependence of magnetization at H=500 Oe field.}
	\label{VMHT}
\end{figure}

\subsection{Magnetic Properties}
For an ordered quaternary Heusler alloy XX$'$YZ, when X$'$ and Y have less than half-filled d-orbitals and Z = Al or Si, the magnetization obeys the so-called Slater-Pauling (SP) rule,\cite{doi:10.1063/1.4805063} according to which the net magnetization of the compound is given by,
\begin{equation}
M = (N_v - 18)\ \mu_B/f.u.
\end{equation}
where N$_v$ is the total number of valence electrons in the alloy.
NbVTiAl has 17 valence electrons and hence expected to have a magnetic moment of 1 $\mu_B/f.u.$, as per the SP rule. Figure \ref{VMHT}(a) shows the M-H curve for NbVTiAl measured at 5 K. It clearly indicate a very small net magnetization ($\sim 10^{-3} \mu_B/f.u.$). However, the initial slope of the M-H loop clearly indicates ferro/ferri magnetic nature of the alloy. These observations along with the M-T data (discussed below) are indicative of nearly compensated ferrimagnetism in this alloy. It should be noted that in the case of an antiferromagnet, the compensated magnetic moment is due to the same kind of ions present at different sites, whereas, FCF materials usually contain three or more magnetic ions with their moments aligned in such a way that the total magnetization is zero. This claim, however, can only be established using atom projected magnetic states determined using advanced experimental probes such as X-ray magnetic circular dichroism (XMCD) or neutron diffraction, which is beyond the scope of the present work. However, in order to get a better insight, we have carried out theoretical simulations for various magnetic configurations in Y-type ordered structures as well as for random disordered structures. We showed that this deviation in moment possibly arise due to the random disorder (see Section \ref{theo} for detailed analysis of theoretical results).

Figure \ref{VMHT}(b) shows the thermo-magnetic (M-T) curve (measured in ZFC mode) in the temperature range 5-1000 K recorded under an applied field of H = 500 Oe. This curve clearly indicates a high magnetic ordering temperature ($>$ 800 K).
\subsection{Transport Properties}
\subsubsection{Longitudinal Resistivity}
\begin{figure}[hbt]
\centering
\includegraphics[width=0.8\linewidth]{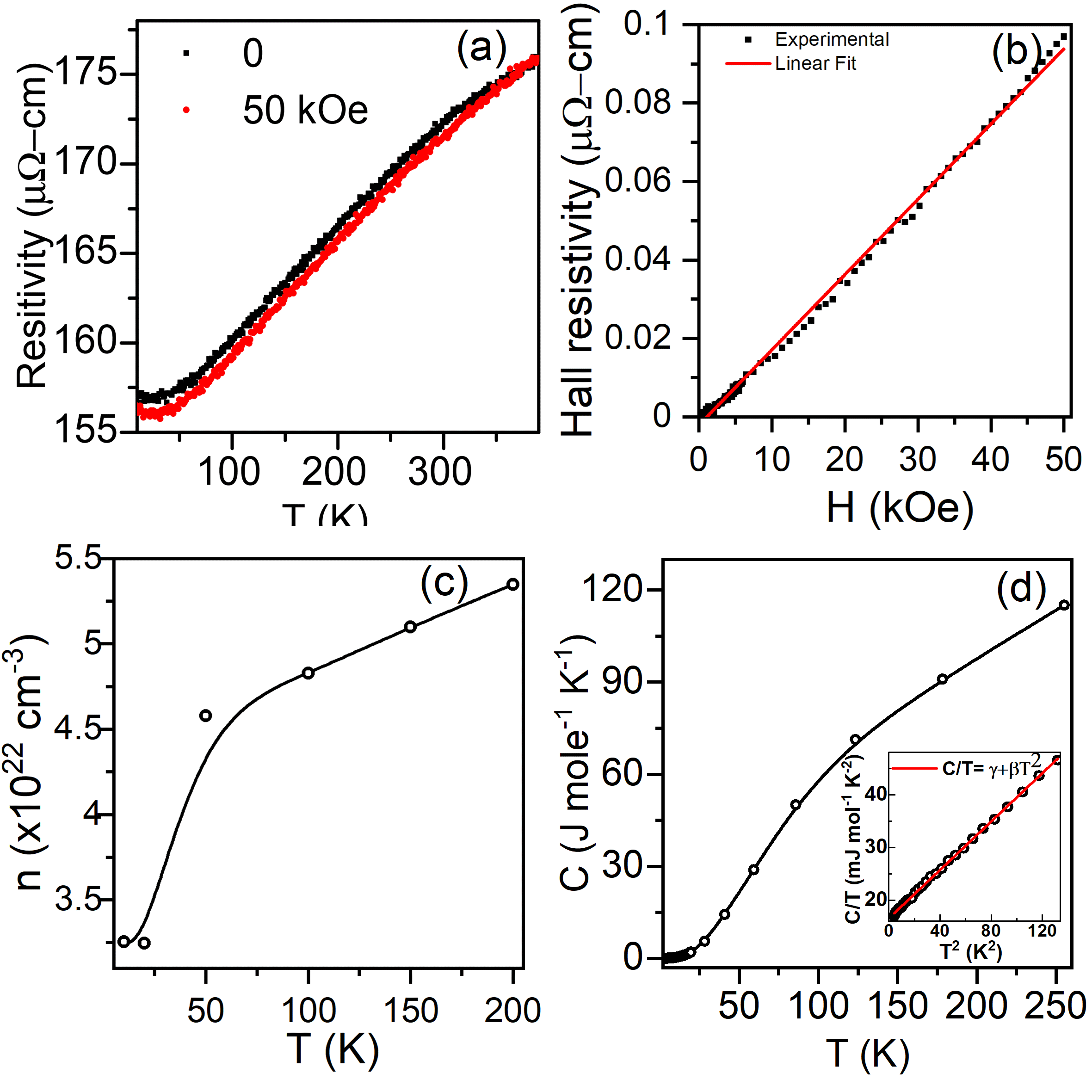}
\caption{(a) Temperature dependence of longitudinal electrical resistivity($\rho$) in 0 and 50 kOe fields, (b) Hall resistivity ($\rho_{xy}$) vs. magnetic field (H), measured at 10 K, along with the linear fit (c) Temperature dependence of carrier concentration (n), and (d) Heat capacity vs. T curve for high T range, up to 250 K. Inset of (d) shows the (C/T) vs.$\mathrm{T^2}$ plot, in low temperature range for NbVTiAl.}
\label{RHC}
\end{figure}

{\color{black}{Figure \ref{RHC}(a) shows the temperature (T) dependence of the longitudinal electrical resistivity of NbVTiAl at 0 and 50 kOe. In the moderately high T range, resistivity increases with temperature revealing metallic nature with a very small negative MR. In case of a metal, the resistivity is generally dominated by electron-phonon scattering which accounts for the linear dependence at high temperatures and saturation near residual resistivity at low temperatures. In the lower temperature region, T $<$ 25 K, the resistivity is almost constant and thus is independent of temperature. A similar behavior has also been observed in other half-metallic Heusler alloys such as CoRhMnGe,\cite{PhysRevB.96.184404} CoRuFeSi,\cite{BAINSLA2015631} Co$_2$Fe$_{1-x}$Cr$_x$Si\cite{PhysRevApplied.10.054022}, Co$_2$FeSi\cite{PhysRevLett.110.066601} etc. and was attributed to the absence of spin-flip scattering. It should be noted that, due to the presence of gap in the minority sub-band, spin-flip scattering is not possible in half-metals. For T $>$ 50 K, the resistivity increases  almost linearly with T, which arises from the electron-phonon scattering.}}

\subsubsection{Hall Effect}
Figure \ref{RHC}(b)  shows the Hall resistivity $\rho_{xy}$ vs. field (H) at 10 K for NbVTiAl. Notably,  $\rho_{xy}$ increases almost linearly with the field. The positive slope of the $\rho_{xy}$ versus H curve indicates that holes are the majority charge carriers. The carrier concentration 'n' was calculated from the slope ($R_H$)  {of $\rho_{xy}$ versus H curves measured at different temperatures using $R_H = 1/(ne)$. Figure \ref{RHC}(c) shows the variation of n with temperature. Clearly, the order of magnitude of carrier concentration is $\sim 10^{22}$ $cm^{-3}$.

\subsubsection{Heat capacity measurements}
In the low temperature range, the heat capacity of a ferromagnetic material with high ${T_C}$ can be described by the Sommerfeld model\cite{PhysRevB.83.235211}(magnetic excitations have insignificant contribution), according to which:
\begin{equation}
C(T)=\gamma T + \beta T^3
\label{CT}
\end{equation}
where, the first and second terms corresponds to the electronic and lattice contributions to the total heat capacity respectively. Here, $\gamma$ is the Sommerfeld coefficient representing the electronic part and $\beta$ is the lattice coefficient. Figure \ref{RHC}(d) shows the C vs. T plot in zero field, while the inset shows the $T^2$ dependence of measured heat capacity (C/T). Evidently, C/T vs. $\mathrm{T^2}$ shows a linear behavior, and the slope and the intercept of the curve correspond to the value of $\gamma$ and $\beta$ of Eq. (\ref{CT}) respectively. In the free electron model, the value of $\gamma$ corresponds to the density of states (DoS) at the Fermi level ($E_F$) according to the relation \cite{met7100414}

\begin{equation}
	N(E_F)=\frac{3\gamma}{\pi^2 N_A {k_B}^2}
\end{equation} 
where, $N_A$ is the Avogadro number and $k_B$ is the Boltzmann constant.

Also, from the value of $\mathrm{\beta}$, the Debye temperature can be calculated using the following relation\cite{met7100414}
\begin{equation}
	\theta_D = \frac{234ZR}{\beta}
\end{equation}
where, $R$ is the universal gas constant and $Z$ is the number of atoms per formula unit.\\

A fitting of C/T vs. $T^2$ curve with Eq. (\ref{CT}) gives $\gamma$ = 16.54 $\mathrm{m J\ mole^{-1} K^{-2}}$ and $\beta$ = 0.23 $\mathrm{m J\ mole^{-1} K^{-4}}$. The calculated value of density of states $N(E_F)$ from the extracted Sommerfeld constant ($\gamma$) is found to be 7.0 states $\mathrm{eV^{-1} f.u.^{-1}}$, while $\theta_D$ is found to be 323 K. It should be noted that, these estimates of $N(E_F)$ and $\theta_D$ are completely based on the free electron model and they only guide us to get a qualitative trend of these numbers as compared to other similar HMFs.

\subsection{Theoretical Results}
\label{theo}
As already mentioned, quaternary Heusler alloys generally crystallise in Y-type structure and there are three possible non-degenerate configurations (namely, Type I, Type II and Type III) depending on the atomic positions of constituent atoms. Total energy ab-initio calculations revealed that Type II configuration in the ferrimagnetic state is energetically the most stable configuration with a net magnetic moment of 0.80 $\mathrm{\mu_B}$/f.u. (The structural optimisation details of the ordered Y-type structure of NbVTiAl can be found in the supplementary material\cite{RR3}.) Also, the calculated spin-polarized band structure and DoS for the energetically most stable, Type II configuration of NbVTiAl alloy confirms a nearly half metallic state with high spin polarization.(See Fig. S1 of supplementary material \cite{RR3}.)

\begin{figure}[hbt]
\centering
\includegraphics[width=0.8\linewidth]{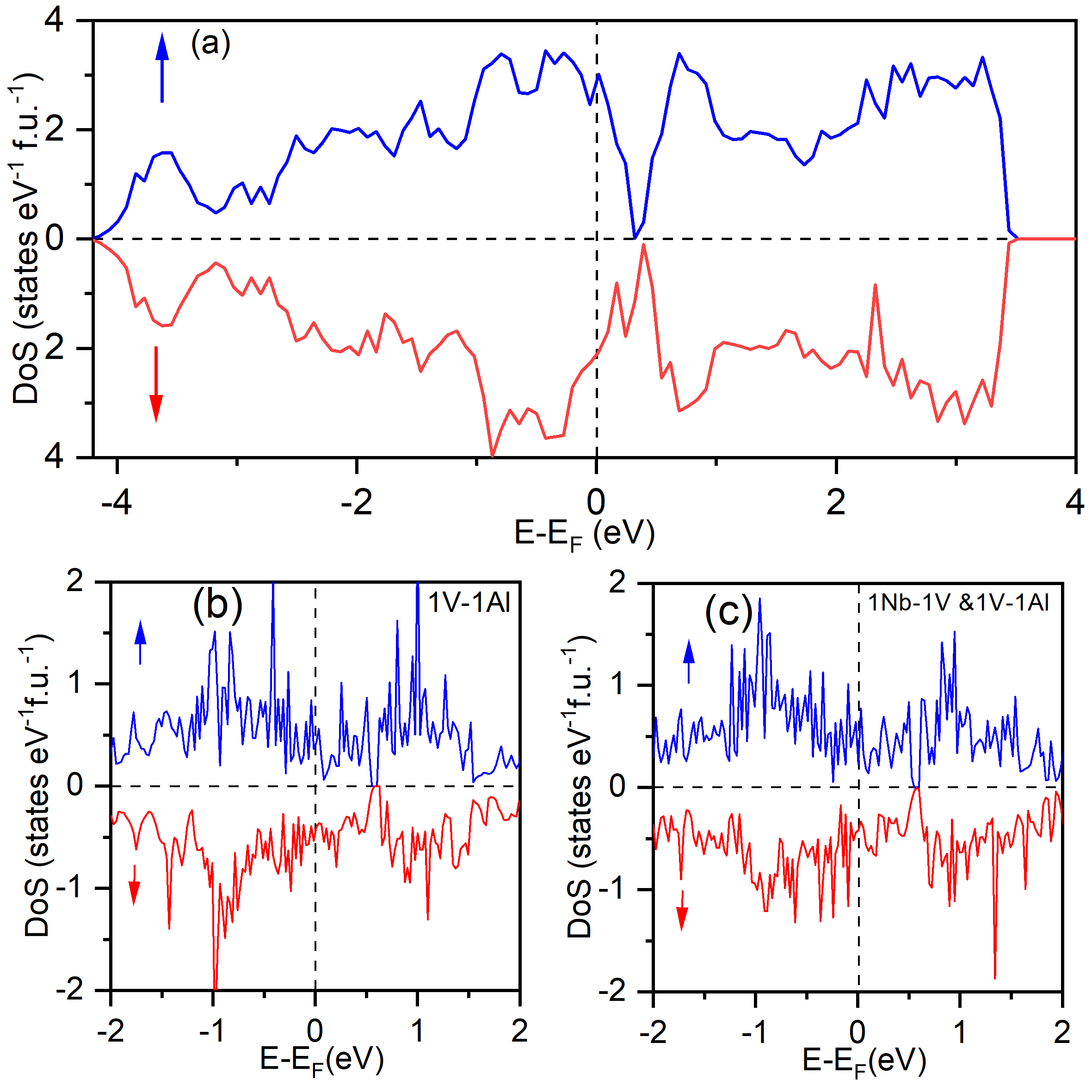}
\caption{Spin polarized density of states for (a) completely disordered A2 structure, and two swap disordered configurations namely (a) 1V-1Al and (b) 1Nb-1V \& 1V-1Al of NbVTiAl. }
\label{dos_dis}
\end{figure}
\begin{table*}[hbt]
	\centering
\caption{Theoretically optimized lattice constants (in \AA), total energy (E) and total magnetic moment of various swap disordered and A2 disordered (SQS) configurations of NbVTiAl . Results for the purely ordered structure is also given for comparison. }
	\begin{tabular}{c| c| c| c}
		\hline \hline
		System & Lattice constants (a/b/c) & E (eV/atom) &$\mathrm{m_{total}}$($\mu_B$/f.u.) \\
		\hline
		Completely Ordered & 6.35\ /6.35\ /6.35 &  -7.874 & 0.80\\
		\hline
		1Nb-1Al  & 6.35\ /6.35\ /6.35 &  -7.869 & 0.50 \\
		2Nb-2Al& 6.37\ /6.37\ /6.37 &-7.852   &0.28\\
		\hline
		1Ti-1Al  & 6.35\ /6.35\ /6.35 & -7.846 & 0.86\\
		2Ti-2Al& 6.38\ /6.35\ /6.35 &-7.817  &0.90\\
		\hline
		1V-1Al  & 6.34\ /6.34\ /6.34 & 7.849& 0.02\\
		2V-2Al& 6.33\ /6.33\ /6.37 & 7.816 &-0.10\\
		\hline
		1Nb-1Ti  & 6.36\ /6.36\ /6.36 & -7.853  & 0.99\\
		2Nb-2Ti& 6.36\ /6.36\ /6.36 & -7.831&0.88\\
		\hline
		1Nb-1V  & 6.35\ /6.35\ /6.35 &-7.863 & 0.64\\
		2Nb-2V& 6.35\ /6.35\ /6.37& -7.855&0.50\\
		\hline
		1V-1Ti  & 6.35\ /6.35\ /6.35 &-7.871  & 0.51\\
		2V-2Ti& 6.35\ /6.35\ /6.35& -7.865 &0.43\\
		\hline
		1Nb-1Al \& 1Ti-1Al  & 6.34\ /6.36\ /6.36 &-7.838 & 0.61\\
		\hline
		1Nb-1V \& 1V-1Al  & 6.36\ /6.36\ /6.35 & -7.841 & 0.01\\
		\hline
		1Nb-1Ti \& 1V-1Al  & 6.36\ /6.36\ /6.35 & -7.833& 0.03\\
		\hline
		A2 disorder & 6.36\ /6.36\ /6.35 & -7.837& 0.05\\
		\hline \hline
	\end{tabular}
	\label{MM}
\end{table*}  
The simulated net magnetic moment (0.80 $\mathrm{\mu_B}$/f.u.) for ordered NbVTiAl agrees fairly well with the Slater-Pauling rule (1.0 $\mathrm{\mu_B}$/f.u.) but goes in complete contrast with the experimental observation where the alloy showed almost negligible ($\sim 10^{-3} \mu_B/f.u.$) net moment.
Such a discrepancy may arise due to several reasons such as long range magnetic ordering, structural disorder \cite{RANI2019165662, PhysRevB.60.13237, doi:10.1063/1.4998308, FENG20157, PhysRevB.66.104429} etc. We have indeed checked several magnetic orders including antiferromagnetic (AFM), 2$^{nd}$ neighbour ferrimagnetic etc., but all of them are either much higher in energy as compared to the ferrimagnetic Type II configuration, or of competing energy but with finite net moment or non-metallic electronic structure, in complete disagreement with the experimental findings. As such, these calculations could not reconcile the discrepancy. The other option left was to explore the effect of substitutional disorder on the electronic and magnetic properties. As demonstrated in Sec. \ref{result1}, our XRD study revealed a A2 type disorder in NbVTiAl i.e., all the four Wyckoff positions are equally probable to be randomly occupied by four atoms Nb, V, Ti and Al.

 To understand the experimental observations better, we dig deep into the local atomic arrangement of the system and discuss its relevance  to the observed net magnetic moment. To do this, we generated a completely disordered A2 structure of NbVTiAl and simulated the same to analyze the repercussions associated with it, on the system's magnetic properties. Interestingly, this structure yields an almost negligible ($\sim$0.05 $\mu_B$/f.u.) net moment for NbVTiAl.
This A2 phase remains energetically very close to the ordered Type II configuration (within 37 meV/atom) (see Table \ref{MM}). It is expected that due to disorder, new states can emerge near the Fermi level, which can further reduces the spin polarization. Interestingly, the A2 disordered phase of NbVTiAl indeed follows this expectation and correctly confirms a metallic phase for NbVTiAl, as observed experimentally. This is clear from the spin-polarized density of states, as shown in Fig. \ref{dos_dis}(a). Also, the calculated value of total density of states $N(E_F)$ for the SQS structure was found to be $\sim$ 4.90 states $\mathrm{eV^{-1} f.u.^{-1}}$ which is comparable to the value ($\sim$7.0 states $\mathrm{eV^{-1} f.u.^{-1}}$) obtained from the extracted Sommerfeld constant (using heat capacity measurements).

To get a better insight into the actual reason for negligible moment for A2 disordered case, we first look at the moments of the individual atoms. We found that there are a few V atoms with moment as high as $\sim$0.4 $\mu_B$, but there are also nearby atomic sites involving V, Ti and Nb which gives small compensating magnetic moments. Overall, it can be inferred that a completely disordered environment can effectively induce randomly oriented magnetic moments which, when averaged out, collectively yields zero net moment. Though it can be concluded that a complete random disorder statistically explains the vanishing net moment, we try to provide more material specific discussion in the following paragraph, as the percentage of disorder might be different in the actual, synthesized sample. In other words, we try to understand the effect of neighboring environment on the local moment of each individual atom.

	NbVTiAl possesses a crystal structure with four interpenetrating fcc lattices. In a completely ordered structure, nearest neighbour environment of V involves 4 Al and 4 Nb atoms. Similarly, Nb has 4 V and 4 Ti atoms in its immediate environment, and so on. In a completely disordered structure, this nearest neighboring environment changes completely, which can make a huge impact on the atom projected local moments and hence the net moment. To get a clear picture of the effect of small change of atomic environment on the local magnetic arrangement in NbVTiAl, we have simulated a few simple point defects introduced in an ordered environment. A $2\times2\times2$ supercell of the primitive cell of the most stable ordered configuration (Type II) of NbVTiAl was constructed. This supercell contains a total of 32 atoms, including 8-atoms of each kind. Several swap-disordered (where one atom was exchanged by another atom) configurations have been simulated to check if such a disorder is responsible for the almost negligibly small measured moment in this alloy. In a $2\times2\times2$ supercell, exchanging one of the eight 'P' atom positions with one of the eight 'Q' atom positions leads to a 12.5\% swap disorder between 'P' and 'Q' atoms. All possible configurations for replacement of 'P' by 'Q' and vice versa were simulated, and the results of energetically most stable configurations are chosen to present here. When the position of one Nb atom was interchanged with 1 Al atom, the disorder configuration was labelled as "1Nb-1Al". Similarly, "2Nb-2Al" type disorder indicates that the positions of two Nb atoms were interchanged with two Al atoms. Various disordered structures have been considered and labeled in a similar manner as described above. Each of these structures was fully relaxed. Table \ref{MM} summarises the optimized lattice parameters, total energies and the net magnetic moments for various simulated disordered structures. It can be seen that all the disordered states, except those by involving swap disorder between Nb and Ti atoms, results in reduced moment as compared to that for a purely ordered structure. In fact, in the case of Nb-Ti swap disordered structures, the total magnetic moment increases as compared to the ordered structure. In some cases, the reduced moment was accompanied by a cubic to pseudo cubic distortion (lattice constants a, b, c differ by a small amount). Almost zero magnetic moment was observed in three cases, 1V-1Al, 1Nb-1V \& 1V-1Al, and 1Nb-1Ti \& 1V-1Al. As compared to the completely ordered structure, the atom projected local moments in the disordered case changes considerably (some even get quenched and/or antiferromagnetically aligned) due to the change in the local atomic environment. Thus, the reduced moment observed in NbVTiAl possibly originates from some specific swap disorder configurations between the constituent atoms, as discussed above.

       To further understand the effect of disorder on the electronic structure of a few swapped disordered configurations having negligibly small total magnetic moments, we have simulated the spin-polarized density of 1V-1Al, 1Nb-1V \& 1V-1Al cases. These are shown in Fig. \ref{dos_dis}(b) and \ref{dos_dis}(c). One can notice that due to disorder, new states emerge near the Fermi level, which make the system metallic, and reducing the spin polarization. Overall, this study provides a clear understanding on how the degree of disorder and hence the altered local atomic environment can affect the net moment and electronic structure of quaternary Heusler alloys.

\section{Conclusion}
\label{conc}
In summary, we report an interesting quaternary alloy NbVTiAl which is observed to crystallize in A2-type (Im-3m) structure. Disorder mainly arises because of the similar electronegativity of the constituent elements. Magnetization measurements confirm a high Curie temperature ($>$ 800 K) and a negligibly small magnetic moment ($\sim 10^{-3} \mu_B/f.u.$) at 5 K. Temperature dependence of longitudinal resistivity reveals a metallic nature. Holes act as majority carries with a carrier concentration of $\sim$ $10^{22}$ cm$^{-3}$, as confirmed by the Hall measurements. Band structure calculation of purely ordered NbVTiAl confirms a nearly half-metallic behavior with a high spin polarization ($\sim$ 90\%). The simulated net magnetization of the purely ordered phase is 0.8 $\mu_B/f.u.$, which is in fair agreement with the SP rule, but deviates significantly from the experimentally measured magnetization. The detailed study confirms that disorder changes the magnetic moment remarkably, leading to a very small value as observed experimentally. Such a disorder is also responsible for the emergence of states at the Fermi level which impacts the half metallicity  and hence the spin polarization of NbVTiAl. Thus, it has been observed that even if theoretical simulation reveals high spin polarization for the ordered structure, the magnetic and electronic properties obtained from experiment may differ completely. NbVTiAl is one of a kind in the quaternary alloys family where disorder plays a decisive role in controlling the magnetic/electronic properties of the alloy.
\section*{Acknowledgment}
DR acknowledges Prof. K. Hono for allowing to use the high-temperature magnetic measurements facility at NIMS, Japan. JK acknowledge the financial support provided by IIT Bombay. AA acknowledges DST- SERB (Grant No. CRG/2019/002050) for funding to support this research.

\section*{References}
\bibliography{bib}

\end{document}